\documentclass{article}

\usepackage[preprint]{neurips_2026}

\usepackage[utf8]{inputenc} 
\usepackage[T1]{fontenc}    
\usepackage{hyperref}       
\usepackage{url}            
\usepackage{booktabs}       
\usepackage{amsfonts}       
\usepackage{amsmath}        
\usepackage{nicefrac}       
\usepackage{microtype}      
\usepackage{xcolor}         
\usepackage{graphicx}       
\graphicspath{{../figures/}} 

\workshoptitle{Sim2Science}

\title{A Missing Latent, Not a Missing Simulator: Radius-Augmented Inference for Real JWST Retrieval}

\author{
  Angshuman Chakravertty\\
  SVKM's NMIMS (Deemed to be University)\\
  \texttt{angshumanchakravertty2@gmail.com}
  \And
  P V V RAJ\\
  SVKM's NMIMS (Deemed to be University)\\
  \texttt{pvvraj1234433@gmail.com}
}

\begin{document}

\maketitle

\begin{abstract}

Amortized simulation-based inference (SBI), which is trained on radiative-transfer simulators, recovers exoplanet atmospheres accurately on synthetic James Webb Space Telescope (JWST) spectra but collapses when it comes to real reduced spectra. The flow posterior collapsed on real WASP-39b (importance-sampling effective sample size (ESS) = 1, best-fit $\chi^2/\mathrm{N}=301$), and such a failure is usually blamed on missing the forward-model physics, but ruling these levers out with nested sampling first (a temperature gradient, $SO_2$ opacity, and a high-fidelity opacity set) leaves the fit unchanged, meaning the collapse is not from the simulator misspecification but instead from a missing latent, the planet radius. To fix this, we build MIRAGE, a radius-augmented flow-matching posterior calibrated against an independent nested-sampling reference with importance sampling and an optimal-transport map, which yields a physical and literature-consistent retrieval of real WASP-39b (with $\chi^2/\mathrm{N}$ from 301 to 0.06). This same method transfers unchanged across two instruments and three real JWST targets, including one spectrum self-reduced end-to-end from raw Mikulski Archive for Space Telescopes (MAST) data. The lesson is cross-domain, as a latent the simulator encodes but the inference omits can masquerade as misspecification.

\end{abstract}

\section{Introduction}
\label{sec:intro}

Currently, amortized simulation-based inference (SBI) and neural posterior estimation (NPE) are the standard for scientific inverse problems where we train a conditional density estimator on simulator draws and get near-instant posteriors at the test time~\citep{cranmer2020frontier}. These estimators are only as trustworthy as the simulator, and on real data they can fail silently: when the simulator is misspecified relative to the instrument, the posterior concentrates confidently in the wrong place~\citep{hermans2022crisis}. James Webb Space Telescope (JWST) transmission spectroscopy is a case where ML retrievers trained on radiative-transfer forward models recover parameters accurately on synthetic spectra but degrade on real reduced spectra~\citep{vasist2023npe, gebhard2025flow}. In this study we use three real JWST targets with an established forward model and validate against nested sampling; we do not claim any new forward model, and the lesson of a missing latent to identifiability collapse is simulator-agnostic.

When an SBI retriever fails on real data, it is usually blamed on \textbf{forward-model fidelity}, like missing molecules, opacity tables, etc., but \textit{is the real-data collapse a fidelity problem or an identifiability problem?} On real WASP-39b the flow posterior collapsed with importance-sampling ESS $=1$ and best-fit $\chi^2/N = 301$, yet candidate physics levers ruled out nested-sampling-first each left the fit unchanged (Section~\ref{sec:diagnosis}). Without the planet radius among the inferred parameters, the network could not absorb a $\sim$3\% transit-baseline offset, causing it to force the atmosphere to an over-confident, wrong solution as seen in Section~\ref{sec:diagnosis}. The failure looks like a model error but is non-identifiability: a single latent that a simulator encodes but the inference omits can turn a well-specified forward model into a misspecified one. We build MIRAGE (Multi-Instrument Retrieval with Adaptive Generative Estimation), which adds radius to $\theta$ and trains on domain-randomized spectra, after which the overdispersed flow posterior is calibrated against an independent nested-sampling reference via an optimal-transport map and importance sampling, which are explained in detail in Section~\ref{sec:method:cal} and Section~\ref{sec:method:sim}. Finally, we use the same method, unchanged, across three real targets spanning planet class (hot Saturn $\leftrightarrow$ cold sub-Neptune), instrument (NIRSpec PRISM $\leftrightarrow$ NIRISS SOSS), and provenance (published $\leftrightarrow$ self-reduced).

We make three contributions. (1) We diagnose the real-data collapse of an SBI JWST retriever as an identifiability failure from a missing radius latent rather than forward-model misspecification, established by ruling candidate physics levers out nested-sampling-first (Section 3). (2) We build a radius-augmented flow-matching posterior, calibrated with optimal transport against a nested-sampling anchor, that yields a physical and literature-consistent retrieval of real WASP-39b with $\chi^2/N$ falling from 301 to 0.06 and coverage rising from 2/7 to 7/7 (Sections 2 and 4). (3) We show the same method transfers unchanged across three real JWST targets and two instruments, including one spectrum self-reduced end-to-end from raw MAST (Section 4). Section 5 gives an honest account of where the method stops, and we will release reproducible artifacts on acceptance.

\section{Method}
\label{sec:method}

\begin{figure}[h]
  \centering
  \includegraphics[width=\linewidth]{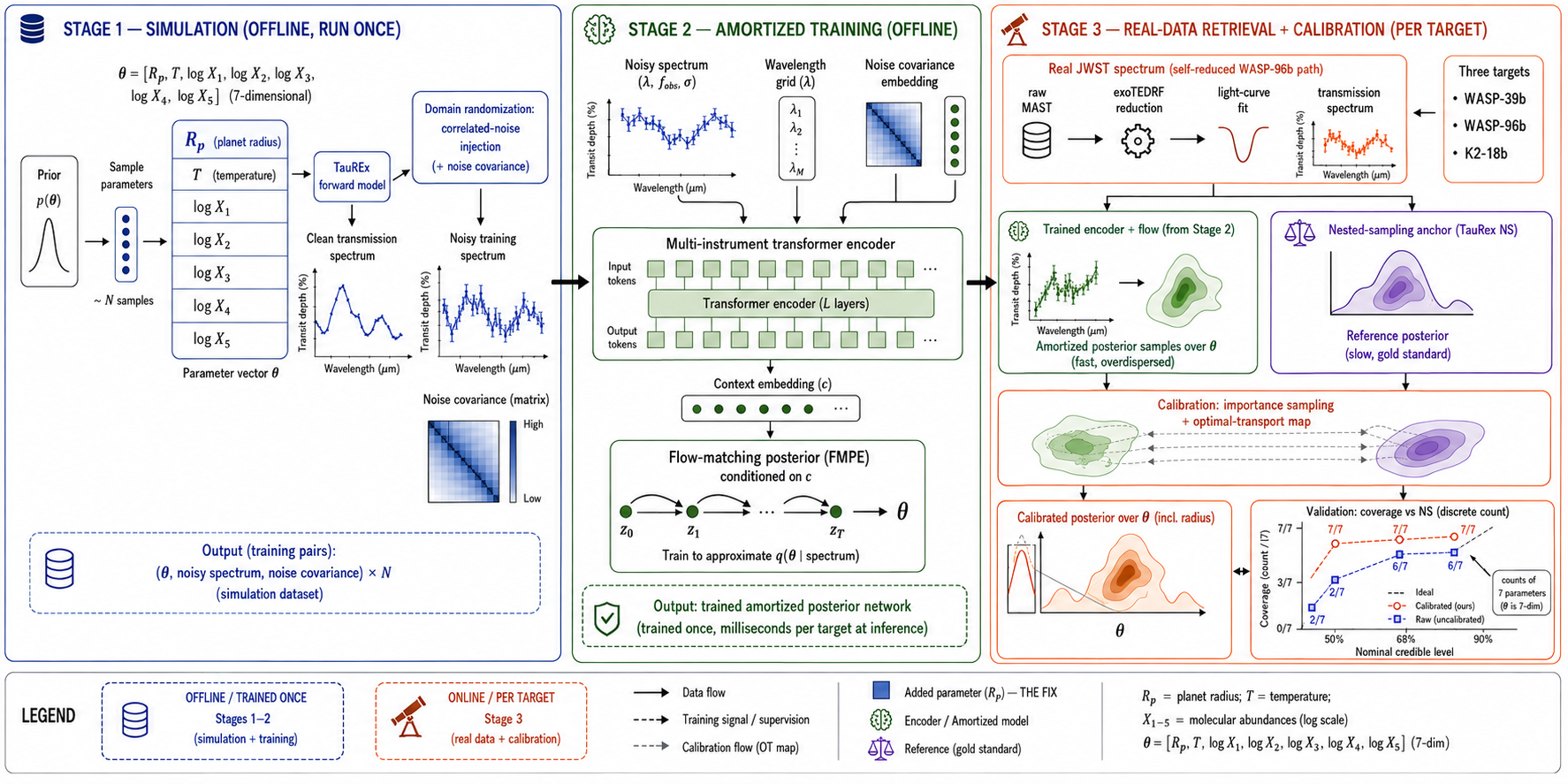}
  \caption{MIRAGE architecture. Stages 1 and 2 (simulation and amortized training) run offline, and Stage 3 retrieves on a real JWST spectrum per target and calibrates the flow posterior against an independent nested-sampling reference.}
  \label{fig:arch}
\end{figure}

\subsection{Forward model and training data}
\label{sec:method:sim}

Training spectra are generated with TauREx 3, a fast, extendable atmospheric retrieval and forward-modelling framework \citep{alrefaie2021taurex}, which maps a vector of atmospheric parameters to a transmission spectrum transit depth as a function of wavelength. We adopt a seven-dimensional parameterization $\theta = [R_p, T, \log X_1, \dots, \log X_5]$: planet radius, an isothermal atmospheric temperature, and the log-abundances of five molecular species. Prior ranges for each parameter are held fixed across training and are reported in full in Appendix~A. Opacity cross-sections are drawn from the ExoMolOP database \citep{chubb2021exomolop}. Spectra are computed on the wavelength grid of the real JWST observations against which the model is ultimately evaluated, so that no interpolation or regridding step is introduced between simulated training data and real test data. This choice keeps the input representation identical across the simulation-to-real boundary, isolating noise and calibration mismatch, not wavelength-grid mismatch, as the object of study in the remainder of this section. The corresponding architecture stage is shown in Figure~\ref{fig:arch} (Stage 1).

A network trained on clean or independently and identically distributed (i.i.d.) noise collapses when applied to real JWST spectra because real instrument systematics are correlated across wavelength rather than i.i.d., a mismatch documented independently across the retrieval literature (see Section~\ref{sec:diagnosis}). We address this at the level of the training distribution itself: rather than training on idealized simulator output, we domain-randomize every training spectrum with structured, correlated noise, so the encoder is exposed during training to the same class of systematics it will encounter on real data. This follows the domain randomization strategy established for simulation-to-real transfer in robotics~\citep{tobin2017domain}, adapted here to spectroscopic time series. The observation model is $$ x = f_{\mathrm{TauREx}}(\theta) + \varepsilon, \qquad \varepsilon \sim \mathcal{N}\big(0,\ \Sigma(\phi)\big), $$ where $\Sigma(\phi)$ is a correlated noise covariance whose hyperparameters $\phi$ are randomized at training time. Concretely, the covariance is \[ \Sigma_{ij} = (1-\rho)\sigma^2\delta_{ij} + \frac{\rho\sigma^2}{2} \exp\!\left[-\frac{1}{2}\left(\frac{|d_{ij}|}{\ell_{\mathrm{SE}}}\right)^2\right] + \frac{\rho\sigma^2}{2} \exp\!\left[-\frac{|d_{ij}|}{\ell_{\mathrm{OU}}}\right], \] where $d_{ij}=\lambda_i-\lambda_j$. A fraction $\rho$ of the total per-bin variance is assigned to correlated structure, split equally between a squared-exponential (SE) component for smooth, detector-drift-like correlations and an Ornstein--Uhlenbeck (OU) component for longer-range, $1/f$-like correlations, while the remaining $(1-\rho)$ is white noise. At each training draw, $\sigma\sim\mathcal{U}(0.05,0.30)$, $\rho\sim\mathcal{U}(0.3,0.8)$, $\ell_{\mathrm{SE}}\sim\mathcal{U}(0.02,1.00)\,\mu\mathrm{m}$, and $\ell_{\mathrm{OU}}\sim\mathcal{U}(0.05,3.00)\,\mu\mathrm{m}$. The resulting covariance is not simply injected into the spectrum; it is also passed to the network as a conditioning input, forming the covariance embedding used in Stage 2. We defer the effect of that conditioning to Sections~\ref{sec:results} and~\ref{sec:limits}, and make no claim about it here. The training set is constructed by sampling $\theta$ from the prior, running the forward model, and injecting a freshly-drawn noise realization, yielding training pairs $(\theta,\ \text{noisy spectrum},\ \Sigma) \times N$, with $N = 28{,}000$. This is the architecture's single point of contact with Figure~\ref{fig:arch} (Stage 1).

\subsection{Flow-matching posterior with radius}
\label{sec:method:flow}

We approximate the posterior with a flow-matching estimator (FMPE)~\citep{wildberger2023fmpe, gebhard2025flow} on the fm4ar backbone, conditioned on the encoder context $c(x)$ from Section~\ref{sec:method:sim}. A Gaussian base is transported by the flow~\citep{lipman2023flow} to the posterior along a linear path, with the conditional flow-matching objective being Equation~\ref{eq:fm}. The learned velocity field is integrated by drawing $\theta$ from Gaussian sample to $t=1$, which is amortized and runs in milliseconds per target.

\begin{equation}
\mathcal{L}(\psi) = \mathbb{E}_{t \sim U[0,1],\; z_0 \sim \mathcal{N}(0, I),\; (\theta, x)}
\Big\| \, v_\psi\big((1-t)\,z_0 + t\,\theta,\; t,\; c(x)\big) - (\theta - z_0) \, \Big\|^2 .
\label{eq:fm}
\end{equation}

In Equation~\ref{eq:theta}, the most notable change is that the inferred vector now incorporates the planet radius. With the inclusion of $r_p$ in $\theta$, the network can now absorb the transit-baseline offset that Section~\ref{sec:diagnosis} identified, instead of distorting the atmosphere to compensate. The gain is attributable to the added latent and not a new architecture because this is the only structural change from a standard FMPE.

\begin{equation}
\theta = [\, r_p,\; T,\; \log X_1,\; \dots,\; \log X_5 \,] \in \mathbb{R}^{7} .
\label{eq:theta}
\end{equation}

\subsection{Calibration}
\label{sec:method:cal}

We reweight the flow proposal $q(\theta\mid x)$ toward the true posterior with importance sampling because on real data it is overdispersed, Equation~\ref{eq:is}. We also monitor the effective sample size $\mathrm{ESS} = 1/\sum_i w_i^2$ after normalizing the weights to sum to one. On a single peaked real likelihood, the raw ESS remains low, thus we report the calibration by coverage rather than by ESS.

\begin{equation}
w_i \;\propto\; \frac{p(x \mid \theta_i)\, p(\theta_i)}{q(\theta_i \mid x)}, \qquad
\theta_i \sim q(\cdot \mid x) .
\label{eq:is}
\end{equation}

The flow proposal is transported onto an independent nested-sampling reference $p_{\mathrm{NS}}$~\citep{alrefaie2021taurex, skilling2006nested} with an optimal-transport map as seen in Equation~\ref{eq:ot}, which was inspired by the optimal-transport treatment of misspecification in RoPE~\citep{wehenkel2025rope}. The calibrated samples are $\tilde\theta_i=T^\star(\theta_i)$, and their quality is measured as coverage of the NS values inside the calibrated intervals and reported in Section~\ref{sec:results}.

\begin{equation}
T^\star \;=\; \arg\min_{T:\; T_\# q \,=\, p_{\mathrm{NS}}} \;
\int \| \theta - T(\theta) \|^2 \, q(\theta \mid x)\, d\theta .
\label{eq:ot}
\end{equation}

\section{Diagnosis: the gap is identifiability, not simulator fidelity}
\label{sec:diagnosis}

On the real WASP-39b, the amortized flow posterior settled on a nearly cold, flat atmosphere, along with the best-fit $\chi^2/\mathrm{N} = 301$ and importance sampling ESS = 1, leaving the posterior confident but wrong. Either the forward model is missing physics, or the inference is missing a parameter. We test the first exhaustively. Additionally, in published retrievals the same class of forward model fits WASP-39b, hinting that the physics is not the issue here but something in the setup differs~\citep{rustamkulov2023wasp39}.

For each candidate missing physics lever, we re-fit it with nested sampling rather than the flow after adding it to the forward model; thus, the test is independent of the amortized network. There are three primary levers with this outcome: a non-isothermal temperature, added SO$_2$ opacity~\citep{tsai2023so2}, and a high-fidelity $R{=}15000$ ExoMolOP opacity set~\citep{chubb2021exomolop}, which are set across all molecules, each added and re-fit without flipping the fit. The cold, flat solution persisted at $\chi^2$ near 2.5 (the unphysical cold-flat corner). As the richer physics was not helpful, we conclude that the collapse is not a forward-model fidelity problem.

\begin{table}[h]
  \centering
  \small
  \caption{Ruling out forward-model fidelity, nested sampling first. Each variant is fit to real WASP-39b with the radius held fixed except the last row and $\chi^2/N$ is measured over the 47 real-covered bins.}
  \label{tab:diagnosis}
  \begin{tabular}{lcc}
    \toprule
    Forward-model variant (radius fixed unless noted) & best-fit $T$ (K) & $\chi^2/N$ \\
    \midrule
    Physical atmosphere, isothermal      & 700 & 38.44 \\
    \quad + temperature gradient         & 700 & 38.48 \\
    \quad + SO$_2$ opacity               & 700 & 38.46 \\
    \midrule
    \textbf{Free planet radius}          & \textbf{606} & \textbf{0.76} \\
    \bottomrule
  \end{tabular}
\end{table}

The planet radius was held fixed, so $\theta$ had no radius term. As the radius trades off against the atmospheric parameters, through the baseline level, a roughly 3\% mismatch between the assumed transit baseline and the real spectrum's continuum could only be absorbed by distorting the atmosphere. When the radius is freed, using the same forward model and data, $\chi^2/\mathrm{N}$ drops from 38.44 to 0.76, and the fit then returns to a warm ($\sim$600 K), water-rich, literature-consistent atmosphere (log H$_2$O $\approx$ -3.2). The failure shows up as model error but is really non-identifiability, which motivated adding radius to $\theta$ in the amortized method (Section~\ref{sec:method:flow}).

\section{Results}
\label{sec:results}

\begin{figure}[h]
  \centering
  \includegraphics[width=\linewidth]{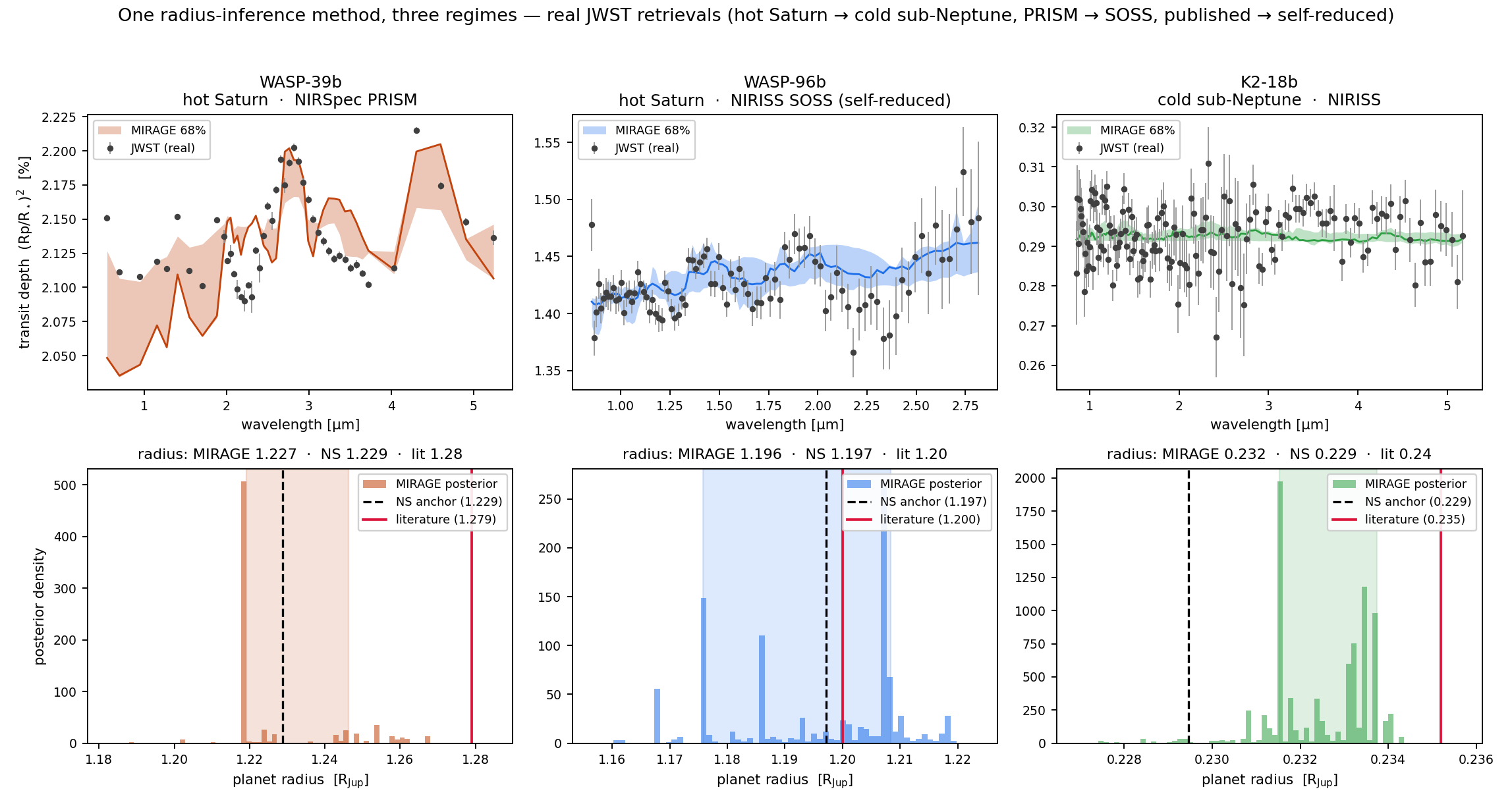}
  \caption{Multi-target retrievals. The same radius-augmented, calibrated method applied unchanged to three real JWST targets that span planet class, instrument, and data provenance. The top row shows the MIRAGE posterior against the real spectrum, and the bottom row shows the recovered planet-radius posterior against the nested-sampling anchor and the literature value.}
  \label{fig:multitarget}
\end{figure}

The radius-augmented posterior gives a physical, literature-consistent fit on real WASP-39b after the OT calibration. There is a drop in the best-fit $\chi^2/\mathrm{N}$ from 301 to 0.06, along with the recovered radius and water abundance matching the nested-sampling anchor and published retrieval, as shown in Figure~\ref{fig:multitarget} and Table~\ref{tab:results}, on the~\citep{rustamkulov2023wasp39} spectrum that produced the Section~\ref{sec:diagnosis} collapse.

Three targets, spanning planet class (hot Saturn $\leftrightarrow$ cold sub-Neptune), instrument (NIRSpec PRISM $\leftrightarrow$ NIRISS SOSS), and provenance (published $\leftrightarrow$ self-reduced), have the same method and are unchanged. The coverage of the NS values inside the calibrated intervals over the seven inferred parameters is 7/7, 7/7, and 6/7 (Table~\ref{tab:results}), with the K2-18b spectrum~\citep{madhusudhan2023k218}. In every case the radius lands near the literature value, and the abundances are broad but physical.

WASP-96b is not a published spectrum but is reduced end-to-end from raw MAST data (NIRISS SOSS)~\citep{radica2024exotedrf} with exoTEDRF and a light-curve~\citep{kreidberg2015batman} fit, and its 1.4 $\mu$m water feature is recovered, making the provenance a real axis of the breadth claim instead of three published spectra. The retrieval on the self-reduced spectrum is as strong as on the published ones.

\begin{table}[t]
  \centering
  \small
  \caption{The same radius-augmented, calibrated method applied unchanged to three real
  JWST targets. Recovered planet radius, with the literature value in parentheses, and the
  coverage of the nested-sampling reference values inside the calibrated credible intervals,
  taken over the seven inferred parameters.}
  \label{tab:results}
  \begin{tabular}{llcc}
    \toprule
    Planet & Instrument & Radius (truth) $R_J$ & Coverage vs NS \\
    \midrule
    WASP-39b & NIRSpec PRISM & 1.227 (1.28)  & 7/7 \\
    WASP-96b & NIRISS SOSS   & 1.196 (1.20)  & 7/7 \\
    K2-18b   & NIRISS SOSS   & 0.232 (0.24) & 6/7 \\
    \bottomrule
  \end{tabular}
\end{table}

As a fair alternative to hand-designed randomization, we also tested whether a learned simulated-to-real domain translation could do this job instead: a CycleGAN is trained to translate simulated spectra toward the distribution of real JWST spectra, using unpaired examples on each side, and the network is then trained on the translated spectra in place of the domain-randomized set \citep{zhu2017cyclegan}. The real side of the unpaired set is drawn from the single reduced WASP-39b spectrum, bootstrap-expanded with 1\% photon-noise jitter to 500 samples; the full 52-bin spectrum is translated, rather than per-instrument segments. The result is a clean negative: translation did not improve on structured randomization. We quantify this with the maximum mean discrepancy (MMD) between translated and target spectra \citep{gretton2012mmd}, which we measure at $\mathrm{MMD} \approx 1.22$. The added generative complexity of a learned translator provides no benefit over the simpler structured-randomization pipeline for this problem, so we will keep the latter as the production training procedure.

\section{Limitations and conclusion}
\label{sec:limits}

Three findings mark the current boundary of the method. First, the amortization gap: calibration currently requires an independent nested-sampling anchor for each new target. The raw flow posterior is overdispersed, and the optimal-transport map, not the flow alone, pulls it onto the reference, so one nested-sampling run is still paid per target. Coverage evaluated against the same reference derives from the construction rather than being independent evidence about the real spectrum since the map is fitted such that the transported samples match the nested-sampling reference. It shows that the transport was solved, not that the network recovered the true posterior. The retrieved radii that match values seen in published values are the portion that is independent of the map. Moving to a higher-resolution training grid improved the network's posterior centering but not its sharpness, which is the same underlying gap surfacing in a different diagnostic \citep{hermans2021crisis}. Second, noise-conditioning does not transfer to real out-of-distribution noise. Covariance conditioning is a clean, reproducible win on the synthetic ABC benchmark, raising coverage from 0.587 to 0.687, with the gain growing as injected noise increases, but this gain does not transfer to a single real target we evaluated, because real instrument noise falls outside the distribution spanned by the training noise model (Section~\ref{sec:method:sim}). We report this as the honest bound it is, rather than as a shortcoming to be explained away. Third, the input-side generative alternative added nothing: the CycleGAN domain translation module tested in Section~\ref{sec:results} did not improve upon structured randomization, reinforcing that the gap being closed here is a distributional one, not one requiring learned generative translation.

A factor the simulator encodes but the inference omits can make even a well-specified forward model look misspecified, so the fix is identifiability, not fidelity. Beyond JWST, this is applicable to any SBI setting where the diagnostic transfers as a recipe, and the proposal omits a factor the simulator represents. Our framework MIRAGE provides a calibrated, physical retrieval of the real JWST spectra across all three targets, which is one added parameter along with calibration, not a new architecture.

\begin{ack}

\end{ack}

\bibliographystyle{plainnat}
\bibliography{references}

@article{cranmer2020frontier,
  author   = {Cranmer, Kyle and Brehmer, Johann and Louppe, Gilles},
  title    = {The frontier of simulation-based inference},
  journal  = {Proceedings of the National Academy of Sciences},
  volume   = {117},
  number   = {48},
  pages    = {30055--30062},
  year     = {2020}
}

@article{hermans2022crisis,
  author   = {Hermans, Joeri and Delaunoy, Arnaud and Rozet, Fran{\c{c}}ois and Wehenkel, Antoine and Louppe, Gilles},
  title    = {A Crisis in Simulation-Based Inference? {B}eware of Miscalibrated Posteriors},
  journal  = {Transactions on Machine Learning Research},
  year     = {2022},
  note     = {Survey Certification}
}

@article{vasist2023npe,
  author   = {Vasist, Malavika and Rozet, Fran{\c{c}}ois and Absil, Olivier and Molli{\`e}re, Paul and Nasedkin, Evert and Louppe, Gilles},
  title    = {Neural posterior estimation for exoplanetary atmospheric retrieval},
  journal  = {Astronomy \& Astrophysics},
  volume   = {672},
  pages    = {A147},
  year     = {2023}
}

@article{chubb2021exomolop,
  author  = {Chubb, Katy L. and Rocchetto, Marco and Yurchenko, Sergei N. and others},
  title   = {The {ExoMolOP} database: Cross sections and $k$-tables for molecules of interest in high-temperature exoplanet atmospheres},
  journal = {Astronomy \& Astrophysics},
  volume  = {646},
  pages   = {A21},
  year    = {2021},
  doi     = {10.1051/0004-6361/202038350}
}

@inproceedings{lipman2023flow,
  author    = {Lipman, Yaron and Chen, Ricky T. Q. and Ben-Hamu, Heli and Nickel, Maximilian and Le, Matt},
  title     = {Flow Matching for Generative Modeling},
  booktitle = {International Conference on Learning Representations (ICLR)},
  year      = {2023}
}

@inproceedings{wildberger2023fmpe,
  author    = {Wildberger, Jonas and Dax, Maximilian and Buchholz, Simon and Green, Stephen R. and Macke, Jakob H. and Sch{\"o}lkopf, Bernhard},
  title     = {Flow Matching for Scalable Simulation-Based Inference},
  booktitle = {Advances in Neural Information Processing Systems (NeurIPS)},
  year      = {2023}
}

@inproceedings{wehenkel2025rope,
  author    = {Wehenkel, Antoine and Gamella, Juan L. and Sener, Ozan and Behrmann, Jens and Sapiro, Guillermo and Cuturi, Marco and Jacobsen, J{\"o}rn-Henrik},
  title     = {Addressing Misspecification in Simulation-based Inference through Data-driven Calibration},
  booktitle = {International Conference on Machine Learning (ICML)},
  series    = {Proceedings of Machine Learning Research},
  volume    = {267},
  year      = {2025}
}

@article{skilling2006nested,
  author  = {Skilling, John},
  title   = {Nested Sampling for General {B}ayesian Computation},
  journal = {Bayesian Analysis},
  volume  = {1},
  number  = {4},
  pages   = {833--860},
  year    = {2006}
}

@article{alrefaie2021taurex,
  author   = {Al-Refaie, Ahmed F. and Changeat, Quentin and Waldmann, Ingo P. and Tinetti, Giovanna},
  title    = {{{TauREx} 3: A Fast, Dynamic, and Extendable Framework for Retrievals}},
  journal  = {The Astrophysical Journal},
  volume   = {917},
  number   = {1},
  pages    = {37},
  year     = {2021},
  doi      = {10.3847/1538-4357/ac0252}
}

@inproceedings{tobin2017domain,
  title     = {Domain randomization for transferring deep neural networks from simulation to the real world},
  author    = {Tobin, Josh and Fong, Rachel and Ray, Alex and Schneider, Jonas and Zaremba, Wojciech and Abbeel, Pieter},
  booktitle = {2017 IEEE/RSJ International Conference on Intelligent Robots and Systems (IROS)},
  pages     = {23--30},
  year      = {2017},
  organization = {IEEE},
  doi       = {10.1109/IROS.2017.8202133}
}

@inproceedings{zhu2017cyclegan,
  title     = {Unpaired Image-to-Image Translation using Cycle-Consistent Adversarial Networks},
  author    = {Zhu, Jun-Yan and Park, Taesung and Isola, Phillip and Efros, Alexei A.},
  booktitle = {Proceedings of the IEEE International Conference on Computer Vision (ICCV)},
  pages     = {2223--2232},
  year      = {2017}
}

@article{gretton2012mmd,
  title     = {A Kernel Two-Sample Test},
  author    = {Gretton, Arthur and Borgwardt, Karsten M. and Rasch, Malte J. and Sch{\"o}lkopf, Bernhard and Smola, Alexander},
  journal   = {Journal of Machine Learning Research},
  volume    = {13},
  number    = {25},
  pages     = {723--773},
  year      = {2012}
}

@article{hermans2021crisis,
  title     = {A Trust Crisis In Simulation-Based Inference? Your Posterior Approximations Can Be Unfaithful},
  author    = {Hermans, Joeri and Delaunoy, Arnaud and Rozet, Fran{\c c}ois and Wehenkel, Antoine and Begy, Volodimir and Louppe, Gilles},
  journal   = {arXiv preprint arXiv:2110.06581},
  year      = {2021}
}

@article{gebhard2025flow,
  title     = {Flow Matching for Atmospheric Retrieval of Exoplanets: Where Reliability meets Adaptive Noise Levels},
  author    = {Gebhard, Timothy D. and Wildberger, Jonas and Dax, Maximilian and Kofler, Annalena and Angerhausen, Daniel and Quanz, Sascha P. and Sch{\"o}lkopf, Bernhard},
  journal   = {Astronomy \& Astrophysics},
  volume    = {693},
  pages     = {A42},
  year      = {2025},
  publisher = {EDP Sciences},
  doi       = {10.1051/0004-6361/202451861}
}

@article{rustamkulov2023wasp39,
  title     = {Early Release Science of the exoplanet {WASP}-39b with {JWST} {NIRSpec} {PRISM}},
  author    = {Rustamkulov, Zafar and Sing, David K. and Mukherjee, Sagnick and May, Erin M. and Kirk, James and Schlawin, Everett and Line, Michael R. and Piaulet, Caroline and Carter, Aarynn L. and Batalha, Natasha E. and others},
  journal   = {Nature},
  volume    = {614},
  number    = {7949},
  pages     = {659--663},
  year      = {2023},
  publisher = {Nature Publishing Group},
  doi       = {10.1038/s41586-022-05677-y}
}

@article{tsai2023so2,
  title     = {Photochemically produced {SO2} in the atmosphere of {WASP}-39b},
  author    = {Tsai, Shang-Min and Lee, Elspeth K. H. and Powell, Diana and Gao, Peter and Zhang, Xi and Moses, Julianne and H{\'e}brard, Eric and Venot, Olivia and Parmentier, Vivien and Jordan, Sean and others},
  journal   = {Nature},
  volume    = {617},
  number    = {7961},
  pages     = {483--487},
  year      = {2023},
  publisher = {Nature Publishing Group},
  doi       = {10.1038/s41586-023-05902-2}
}

@article{madhusudhan2023k218,
  author   = {Madhusudhan, Nikku and Sarkar, Subhajit and Constantinou, Savvas and others},
  title    = {Carbon-bearing Molecules in a Possible Hycean Atmosphere},
  journal  = {The Astrophysical Journal Letters},
  volume   = {956},
  number   = {1},
  pages    = {L13},
  year     = {2023}
}

@article{radica2024exotedrf,
  author   = {Radica, Michael},
  title    = {{exoTEDRF}: An {EXO}planet {T}ransit and {E}clipse {D}ata {R}eduction {F}ramework},
  journal  = {{Journal of Open Source Software}},
  volume   = {9},
  number   = {100},
  pages    = {6898},
  year     = {2024}
}

@article{kreidberg2015batman,
  author   = {Kreidberg, Laura},
  title    = {batman: {BA}sic {T}ransit {M}odel c{A}lculatio{N} in {P}ython},
  journal  = {Publications of the Astronomical Society of the Pacific},
  volume   = {127},
  number   = {957},
  pages    = {1161},
  year     = {2015}
}


\section*{NeurIPS Paper Checklist}

\begin{enumerate}

\item {\bf Claims}

\item[] Question: Do the main claims made in the abstract and introduction accurately reflect the paper's contributions and scope?

\item[] Answer: \textbf{Yes}.

\item[] Justification: The paper's claims are stated in the abstract and introduction, the identifiability diagnosis (Section~\ref{sec:diagnosis}), the radius-augmented calibrated retrieval of WASP-39b with best-fit chi-squared/N from 301 to 0.06 (Section~\ref{sec:results}), along with the transfer across three targets and two instruments (Table~\ref{tab:results}). All are supported by the cited sections, and the scope limits are stated in Section~\ref{sec:limits}

\item {\bf Limitations}

\item[] Question: Does the paper discuss the limitations of the work performed by the authors?

\item[] Answer: \textbf{Yes}.

\item[] Justification: Section~\ref{sec:limits} explicitly discusses the amortization gap, the need for an independent nested-sampling anchor, the failure of covariance conditioning to transfer to real out-of-distribution noise, and the lack of benefit from CycleGAN translation.

\item {\bf Theory assumptions and proofs}

\item[] Question: For each theoretical result, does the paper provide the full set of assumptions and a complete (and correct) proof?

\item[] Answer: \textbf{N/A}.

\item[] Justification: The current paper presents an empirical method and experiments rather than theorem-level theoretical results requiring formal proofs.

\item {\bf Experimental result reproducibility}

\item[] Question: Does the paper fully disclose all the information needed to reproduce the main experimental results of the paper to the extent that it affects the main claims and/or conclusions of the paper (regardless of whether the code and data are provided or not)?

\item[] Answer: \textbf{Yes}.

\item[] Justification: The forward model (TauREx 3, ExoMolOP opacities), the parameterization and priors, the structured-noise model and its hyperparameter ranges, the flow-matching objective (Equation~\ref{eq:fm}), the calibration (Equation~\ref{eq:is} and~\ref{eq:ot}), and the training-set size (N = 28,000) are given in Section~\ref{sec:method} and Appendix A.

\item {\bf Open access to data and code}

\item[] Question: Does the paper provide open access to the data and code, with sufficient instructions to faithfully reproduce the main experimental results, as described in supplemental material?

\item[] Answer: \textbf{No}.

\item[] Justification: The manuscript says that reproducible artifacts are released, but no anonymized code/data URL or concrete reproduction instructions are present in the supplied main.tex. Change this to Yes only after the submission contains an appropriate anonymized access path and instructions.

\item {\bf Experimental setting/details}

\item[] Question: Does the paper specify all the training and test details (e.g., data splits, hyperparameters, how they were chosen, type of optimizer) necessary to understand the results?

\item[] Answer: \textbf{Yes}.

\item[] Justification: Section 2 and Appendix A give the training-data construction, noise model and ranges, the flow-matching objective, the fm4ar backbone, and the optimizer, learning rate, and training schedule.

\item {\bf Experiment statistical significance}

\item[] Question: Does the paper report error bars suitably and correctly defined or other appropriate information about the statistical significance of the experiments?

\item[] Answer: \textbf{No}.

\item[] Justification: The supplied manuscript reports point values such as $\chi^2/N$, coverage counts, and MMD, but does not currently provide error bars, confidence intervals, repeated-run variability, or a statistical-significance procedure for the main experimental comparisons.

\item {\bf Experiments compute resources}

\item[] Question: For each experiment, does the paper provide sufficient information on the computer resources (type of compute workers, memory, time of execution) needed to reproduce the experiments?

\item[] Answer: \textbf{Yes}.

\item[] Justification: All models were trained on a single Apple Silicon machine (MPS, CPU for the covariance arm), not a GPU cluster, as detailed in Appendix A; the models are small (about 1M parameters) and inference is amortized at milliseconds per target.

\item {\bf Code of ethics}

\item[] Question: Does the research conducted in the paper conform, in every respect, with the NeurIPS Code of Ethics?

\item[] Answer: \textbf{Yes}.

\item[] Justification: The work is a scientific-methodology study using simulated spectra and existing JWST observations; the supplied manuscript describes no human-subject experimentation, harmful deployment, or other activity that would indicate a deviation from the NeurIPS Code of Ethics.

\item {\bf Broader impacts}

\item[] Question: Does the paper discuss both potential positive societal impacts and negative societal impacts of the work performed?

\item[] Answer: \textbf{N/A}.

\item[] Justification: This is a methodological study of atmospheric retrieval from public astronomical data, with no direct societal application or foreseeable harmful use.

\item {\bf Safeguards}

\item[] Question: Does the paper describe safeguards that have been put in place for responsible release of data or models that have a high risk for misuse?

\item[] Answer: \textbf{N/A}.

\item[] Justification: The work does not describe a high-risk model, generative system for public deployment, or scraped dataset that creates the safeguard scenario described by this checklist item.

\item {\bf Licenses for existing assets}

\item[] Question: Are the creators or original owners of assets (e.g., code, data, models), used in the paper, properly credited and are the license and terms of use explicitly mentioned and properly respected?

\item[] Answer: \textbf{Yes}.

\item[] Justification: All external assets (TauREx 3, ExoMolOP, exoTEDRF, batman, the fm4ar backbone, and the JWST spectra from MAST) are cited to their source and used under their public licenses.

\item {\bf New assets}

\item[] Question: Are new assets introduced in the paper well documented and is the documentation provided alongside the assets?

\item[] Answer: \textbf{No}.

\item[] Justification: The paper states that reproducible artifacts are released, but the supplied manuscript does not identify the released assets, their documentation, or their license. Update this answer after the anonymized artifact package is finalized.

\item {\bf Crowdsourcing and research with human subjects}

\item[] Question: For crowdsourcing experiments and research with human subjects, does the paper include the full text of instructions given to participants and screenshots, if applicable, as well as details about compensation (if any)?

\item[] Answer: \textbf{N/A}.

\item[] Justification: The supplied paper does not involve crowdsourcing experiments or research participants.

\item {\bf Institutional review board (IRB) approvals or equivalent for research with human subjects}

\item[] Question: Does the paper describe potential risks incurred by study participants, whether such risks were disclosed to the subjects, and whether Institutional Review Board (IRB) approvals (or an equivalent approval/review based on the requirements of your country or institution) were obtained?

\item[] Answer: \textbf{N/A}.

\item[] Justification: The supplied paper does not involve human subjects.

\item {\bf Declaration of LLM usage}

\item[] Question: Does the paper describe the usage of LLMs if it is an important, original, or non-standard component of the core methods in this research?

\item[] Answer: \textbf{N/A}.

\item[] Justification: No LLM is described as part of the scientific method, model, training pipeline, or experimental evaluation in the supplied manuscript. If an LLM was used as an important or non-standard component of the research itself, this answer must be revisited.

\end{enumerate}

\appendix
\section{Training details and priors}
\label{app:details}

\paragraph{Priors.} All parameters use uniform priors, fixed across
training. For the hot-Saturn targets (WASP-39b, WASP-96b) the planet
radius is $R_p\sim\mathcal{U}(0.8,1.6)\,R_J$ and the temperature is
$T\sim\mathcal{U}(300,2500)\,\mathrm{K}$; for the cold sub-Neptune
K2-18b, $R_p\sim\mathcal{U}(0.10,0.40)\,R_J$ and
$T\sim\mathcal{U}(100,600)\,\mathrm{K}$. The five log-abundances use
$\log X_i\sim\mathcal{U}(-9,-1)$ in both regimes. The five species are
H$_2$O, CO$_2$, CH$_4$, CO, NH$_3$.

\paragraph{Network and training.} The context encoder is a transformer of
4 layers, width 128, 4 heads, feed-forward width 512, and dropout 0.1,
producing a 256-dimensional spectrum embedding concatenated with a
64-dimensional residual-covariance embedding (320-dimensional context).
Training uses AdamW at learning rate $3\times10^{-4}$ with a
ReduceLROnPlateau schedule (factor 0.5, patience 10), batch size 512, and
gradient clipping at max-norm 1.0, for up to 512 epochs with early
stopping (patience 50, converging near epoch 250). The set has
$N=28{,}000$ training pairs and 4{,}000 held out for validation.

\paragraph{Compute.} All models train on a single Apple Silicon machine,
the covariance arm on CPU (its batched-covariance operations are unstable
on the MPS backend) and the remaining arms on MPS, rather than on a GPU
cluster. Each model is small (about 1M parameters) and inference is
amortized at milliseconds per target.

\paragraph{Real spectra and nested-sampling anchor.} WASP-39b uses the
NIRSpec PRISM spectrum of \citet{rustamkulov2023wasp39} (52 bins, 47
real-covered used for $\chi^2$); WASP-96b is self-reduced from raw MAST
NIRISS SOSS with exoTEDRF \citep{radica2024exotedrf}; K2-18b uses the
NIRISS spectrum of \citet{madhusudhan2023k218}. The nested-sampling reference
is computed with TauREx using the Nestle sampler with 300 live points.

\end{document}